\documentclass[12pt,notitlepage]{article}
\usepackage[english]{babel}
\usepackage{apacite}
\usepackage{graphicx}
\usepackage{amsfonts}
\usepackage{amssymb}
\usepackage{amsmath}
\usepackage{geometry}
\usepackage{scalefnt}
\usepackage[doublespacing]{setspace}
\usepackage[title,page]{appendix}
\usepackage{float}
\usepackage{apalike}
\usepackage{sectsty}

\input{tcilatex}
\renewcommand{\baselinestretch}{1}
 \sectionfont{\centering}

\begin{document}

\begin{center}
\textbf{Dependent Dirichlet Process Rating Model (DDP-RM)}
\end{center}

\bigskip

\bigskip

\bigskip

\bigskip

\bigskip

\begin{center}
\bigskip

\bigskip

\bigskip

Ken Akira Fujimoto

and

George Karabatsos

University of Illinois-Chicago

\bigskip

\bigskip

March 20, 2013

\bigskip

\bigskip

\bigskip

\bigskip

\bigskip

\bigskip

\bigskip

\bigskip

\bigskip

\bigskip

\bigskip

\bigskip

\bigskip

\bigskip

\bigskip
\end{center}

\bigskip

\bigskip

\bigskip

\bigskip

\bigskip

\bigskip

\bigskip

\bigskip

\bigskip

\bigskip

\bigskip

\bigskip

\underline{Acknowledgements}:\ This research is supported by NSF research
grant SES-1156372, from the Program in Methodology, Measurement, and
Statistics. This paper will be presented at the National Council for
Measurement in Education (NCME)\ Conference, April 26-30, at San Francisco.
We thank Professor Stephen G. Walker, University of Kent, for helpful
conversations about the MCMC\ algorithm we used for the paper. Also, we
thank two anonymous reviewers for their comments.

\newpage

\begin{center}
\textbf{Dependent Dirichlet Process Rating Model (DDP-RM)}

\bigskip

\textbf{Abstract}
\end{center}

\bigskip

Typical IRT\ rating-scale models assume that the rating category threshold
parameters are the same over examinees. However, it can be argued that many
rating data sets violate this assumption. To address this practical
psychometric problem, we introduce a novel, Bayesian nonparametric IRT\
model for rating scale items. The model is an infinite-mixture of Rasch
partial credit models, based on a localized Dependent Dirichlet process
(DDP). The model treats the rating thresholds as the random parameters that
are subject to the mixture, and has (stick-breaking)\ mixture weights that
are covariate-dependent. Thus, the novel model allows the rating category
thresholds to vary flexibly across items and examinees, and allows the
distribution of the category thresholds to vary flexibly as a function of
covariates. We illustrate the new model through the analysis of a simulated
data set, and through the analysis of a real rating data set that is
well-known in the psychometric literature. The model is shown to have better
predictive-fit performance, compared to other commonly used IRT\ rating
models.

\bigskip


\noindent RUNNING TITLE:\ Dependent Dirichlet Process Rating Model.\newline

\bigskip

\bigskip

\bigskip

\bigskip

\bigskip

\bigskip

\bigskip

\noindent

\noindent \pagebreak

\section{Introduction}

In social science research, it is often of interest to analyze examinee
ratings to items of a test. An IRT\ rating model, and its parameters
estimated from the given rating scale data set, provides useful information
about various psychometric qualities. They include the difficulty parameter
of each test item, the thresholds parameters of the rating categories, and
the test ability (latent trait)\ parameter of each examinee. Typical IRT\
rating models include the Rasch rating scale model (Andrich, 1978\nocite%
{andrich1978rating}), partial credit models (PCM)\ (Masters, 1982\nocite%
{masters1982rasch}; Muraki, 1992\nocite{Muraki92}), and the family of graded
response models (Samejima, 1969, 1972\nocite{Samejima72}\nocite{Samejima69}%
), all of which have seen many successful applications in a wide range of
research settings.

Nevertheless, these IRT models have their limitations. Typical IRT\ rating
models assume that the rating category threshold parameters apply to all
examinees. However, this assumption is violated when differential rating
category usage occurs across the examinees. Differential rating category
usage may be caused by differential item functioning (DIF); that is, when
different clusters (groups)\ of examinees give rise to different threshold
estimates for the rating categories after controlling for the level of
examinee ability. The different clusters may either refer to unknown latent
groups or known examinee groups (e.g., male and female). Differential
category usage across examinees may also arise from non-systematic random
error, such as when unclear labels that are assigned to the rating
categories. Regardless, if a typical IRT\ model is used to analyze data
which violates its assumption of no differential category usage, then the
model may poorly fit the data and produce misleading results. The results
would wrongly indicate that, for each test item, a single set of rating
category threshold estimates applies for all examinees. In turn, this could
lead to misleading examinee ability estimates. With traditional models, item
fit statistics are often relied upon to identify items that misfit the
model. However, fit statistics have low power in identifying DIF\ items
(Seol, 1999\nocite{seol1999detecting}; Smith \& Suh, 2003\nocite%
{smith2003rasch}). Moreover, even when an item fit statistic identifies an
item as problematic, it does not explain why the item is misfitting.

Multiple-group IRT\ models (e.g., Lord, 1980\nocite{Lord1980IRT}; Wright \&
Masters, 1982\nocite{WrightMasters82}) are more appropriate when the
differential category usage is a result of DIF. These models specify
interaction covariates between person and item characteristics (e.g.,
overall item difficulty and category thresholds). The regression
coefficients of these interaction terms indicate whether DIF\ is present in
an item, and provide some explanation about how rating category usage varies
as a function of examinee characteristics. This modeling approach, however,
is still limited because it assumes that the model contains all the
covariates that could be associated in explaining DIF. As previously
mentioned, latent or unknown examinee characteristics may also contribute to
differential rating category usage, and/or random error may be present in
the rating thresholds.

It then seems preferable to specify a discrete-mixture IRT\ rating model
that can identify and account for differential rating category usage in the
items, which may either result from multiple latent clusters (groups) of
examinees, and/or result from known examinee characteristics (covariates).
For each item, and conditioned on any other known covariates, the model
would specify a (mixture)\ distribution for the rating category thresholds
over all examinees, while assigning a distinct set of rating category
threshold parameters to each latent cluster of examinees. If all examinees
use (e.g., interpret) an item's rating categories in the same manner, then
the model's threshold distribution becomes unimodal with near-zero variance.
Such a distribution would indicate a single cluster of examinees in terms of
the rating thresholds, as in typical IRT\ rating models which assume no
differential category usage. When an item exhibits differential rating
category usage over examinees, then the model's threshold distribution will
have noticeable variance, with possible skewness and/or multimodality. A
unimodal distribution with noticeable variance and/or skewness may either
indicate uncertainty in the rating category usage of the item or DIF. A
multimodal distribution would indicates DIF, with the multiple modes
indicating multiple latent clusters of examinees. Finally, if an item's
threshold distribution is shown to depend on one or more known covariates
that describe examinee background characteristics (e.g., gender, income),
after controlling for examinee ability, then there is DIF\ due to known
examinee groupings (as in multiple-group IRT).

A discrete mixture model has the general form (e.g., McLachlan \&\ Peel, 2000%
\nocite{mclachlan2000finite}):

\begin{equation*}
f_{G_{\mathbf{x}}}(y|\mathbf{x})=\dint f(y|\mathbf{x};\boldsymbol{\xi },%
\mathbf{\Psi }(\mathbf{x}))\mathrm{d}G_{\mathbf{x}}(\mathbf{\Psi }%
)=\dsum\limits_{h=1}^{H}\,f(y|\mathbf{x};\boldsymbol{\xi },\mathbf{\Psi }%
_{h}(\mathbf{x}))\omega _{h}(\mathbf{x}),
\end{equation*}%
given a mixing distribution $G_{\mathbf{x}}$ that is possibly covariate ($%
\mathbf{x}$) dependent; component indices $h=1,\ldots ,H$, kernel
(component) densities $f(y|\mathbf{x};\boldsymbol{\xi },\mathbf{\Psi }_{h}(%
\mathbf{x}))$ ($h=1,\ldots ,H$) with fixed parameters $\boldsymbol{\xi }$
and random parameters $\mathbf{\Psi }_{h}(\mathbf{x})$ that are subject to
the mixture; and given mixing weights $(\omega _{h}(\mathbf{x}))_{h=1}^{H}$
which sum to 1 at every $\mathbf{x}\in \mathcal{X}$. Mixture IRT\ models
treat $y\in \{k=0,1,\ldots ,m\}$ as a scored item response (e.g., a rating),
and specify each of the kernel densities $f(y|\mathbf{x};\boldsymbol{\xi },%
\mathbf{\Psi }_{h}(\mathbf{x}))$ by an ordinary IRT model, such as a
2-parameter logistic model, or a Rasch rating scale model.

Typical IRT\ mixture models assume finite mixtures (i.e., $H<\infty $)
(Rost, 1991\nocite{rost1991logistic}; Smit, Kelderman, \& van der Flier, 2003%
\nocite{smit2003latent}; Von Davier \& Yamamoto, 2004\nocite%
{von2004partially}; Frick, Strobl, Leisch, \& Zeileis, 2012\nocite%
{frick2012flexible}), which limits their ability to adequately describe many
rating scale data sets. We could achieve greater modeling flexibility by
turning to a fully nonparametric framework, through the specification of an
infinite-mixture model (i.e., $H=\infty $). Such a model has infinitely-many
parameters, and avoids the restrictive assumption of parametric IRT\ models,
namely, that the distribution of item response data can be fully-described
by finitely-many parameters. Along these lines, infinite-mixture IRT\ models
have been developed. They include models based on the Dirichlet process
(DP)\ mixture of the item parameters of a 3-parameter logistic model
(Miyazaki \& Hoshino, 2009\nocite{miyazaki2009bayesian}), models based on a
DP mixture of ability parameters in a Rasch model (San Martin et al., 2011%
\nocite{san2011bayesian}), and a Dependent Dirichlet process (DDP)\ mixture
model for the link function of the 2-parameter IRT\ model (Duncan \&
MacEachern, 2008\nocite{duncan2008nonparametric}). Karabatsos and Walker
(2013 to appear\nocite{KarabatsosWalker2013}) review the DP and DDP mixture\
models for IRT. However, none of the available mixture IRT models provide
clustering of examinees in terms of the rating category threshold
parameters. This is because they do not treat the rating category threshold
parameters as the random parameters (i.e., $\mathbf{\Psi }_{h}(\mathbf{x})$)$%
\ $that are subject to the mixture.

To address the limitations of the existing IRT\ models, we introduce a novel
Bayesian nonparametric IRT\ rating model, which we call the DDP Rating Model
(DDP-RM). This model is an infinite-mixture of Rasch partial credit models,
with rating category threshold parameters subject to the mixture, and with
covariate-dependent stick-breaking weights. The random parameters and the
mixture weights are modeled by a Dependent Dirichlet process (DDP)
(MacEachern, 1999\nocite{MacEachern99}; 2000\nocite{MacEachern00}; 2001%
\nocite{MacEachern01}), which is defined by a novel modification of the
local Dirichlet process (lDP) (Chung \& Dunson, 2011\nocite{ChungDunson11}).

In Section 2, we introduce our DDP Rating Model (DDP-RM). In Section 3, we
illustrate our model on simulated data, in order to demonstrate the model's
ability to identify DIF as a result of latent (unknown) examinee
characteristics (covariates). In Section 4, we illustrate our model on a
real data set of rating scale items, which has been extensively studied in
the psychometric modeling literature (De Boeck \& Wilson, 2004\nocite%
{de2004explanatory}). In this illustration, we also compare the goodness of
predictive fit between our DDP-RM and other IRT\ rating scale model of
common usage. In Section 5, we conclude by discussing possible future
extensions of our model. Throughout, we denote \textrm{n}$\left( \mathbf{%
\cdot }|\mathbf{\cdot },\mathbf{\cdot }\right) $, \textrm{n}$_{p}\left( 
\mathbf{\cdot }|\mathbf{\cdot },\mathbf{\cdot }\right) $, \textrm{ga}$(%
\mathbf{\cdot }|\mathbf{\cdot },\mathbf{\cdot }),$ \textrm{ig}$(\mathbf{%
\cdot }|\mathbf{\cdot },\mathbf{\cdot }),$ \textrm{beta}$(\mathbf{\cdot }|%
\mathbf{\cdot },\mathbf{\cdot })$, and \textrm{un}$(\mathbf{\cdot }|\mathbf{%
\cdot },\mathbf{\cdot })$ as the probability density functions for the
univariate normal, $p$-variate Normal, gamma, inverse gamma, beta, and
uniform distributions, respectively. The gamma and inverse gamma
distributions are parameterized by shape and rate parameters.

\section{The Dependent Dirichlet Process Rating Model (DDP-RM)\label{Section
on DDP-RM}}

Our rating model, the DDP-RM, is defined by an infinite mixture of IRT\
rating model. Specifically, this mixture model assumes that the probability
of a rating $Y=y$ is defined by:\ 
\begin{equation}
P\left( Y=y|\mathbf{x;}\theta ,\boldsymbol{\upsilon },\boldsymbol{\gamma },%
\boldsymbol{\psi }\right) =\dint f\left( y|\theta ,\boldsymbol{\tau }\right) 
\text{\textrm{d}}G_{\mathbf{x}}\left( \boldsymbol{\tau }\right)
=\dsum\limits_{h=1}^{\infty }f\left( y|\theta _{t}\mathbf{,}\boldsymbol{\tau 
}_{h}\right) \omega _{h}\left( \mathbf{x}^{\intercal }\boldsymbol{\gamma };%
\boldsymbol{\upsilon },\boldsymbol{\gamma },\boldsymbol{\psi }\right) ,
\label{Equation: DDP-rm}
\end{equation}%
where the kernel probability densities $f\left( y|\theta \mathbf{,}%
\boldsymbol{\tau }_{h}\right) $ are specified by the partial credit model
(PCM),%
\begin{equation}
f\left( y|\theta \mathbf{,}\boldsymbol{\tau }_{h}\right) =P\left( Y=y|\theta 
\mathbf{,}\boldsymbol{\tau }_{h}\right) =\frac{\exp \left( y\theta
-\sum_{l=0}^{y}\tau _{lh}\right) }{\tsum\nolimits_{k=0}^{m}\exp (k\theta
-\sum_{l=0}^{k}\tau _{lh})},\text{ }h=1,2,\ldots ,
\label{Component densities}
\end{equation}%
where the mixture distribution $G_{\mathbf{x}}$ is covariate ($\mathbf{x}$)
dependent and defined by:%
\begin{equation}
G_{\mathbf{x}}(\cdot )=\dsum\limits_{h=1}^{\infty }\omega _{h}\left( \mathbf{%
x}^{\intercal }\boldsymbol{\gamma }\right) \delta _{\boldsymbol{\tau }%
_{h}\left( \mathbf{x}^{\intercal }\boldsymbol{\gamma }\right) }(\cdot ),
\label{MixDist1}
\end{equation}%
and where $\delta _{\boldsymbol{\tau }}(\cdot )$ denotes the degenerate
distribution which assigns probability 1 to the value $\boldsymbol{\tau }$.
Additionally, $\theta _{t}$ denotes the ability parameter of a given
examinee $t$, for a sample of examinees indexed by $t=1,2,\ldots ,N$; and
for the $m+1$ rating categories indexed by $k=0,1,...,m$, the vector $%
\boldsymbol{\tau }_{h}=\left( \tau _{1h},...,\tau _{mh}\right) ^{\intercal }$
gives the set of rating category threshold parameters for the $h^{\text{th}}$
mixture component, while assuming the constraint $\tau _{0h}\equiv 0$. The
mixture distribution $G_{\mathbf{x}}(\boldsymbol{\tau })$ for the
thresholds, and the corresponding covariate ($\mathbf{x}$)-dependent mixture
weights $\left\{ \omega _{h}\left( \mathbf{x}^{\intercal }\boldsymbol{\gamma 
}\right) \right\} _{h=1,2,...}$\ and atoms $\left\{ \boldsymbol{\tau }%
_{h}\left( \mathbf{x}^{\intercal }\boldsymbol{\gamma }\right) \right\}
_{h=1,2,...}$, are modeled by a modified local Dirichlet Process (lDP)
prior. Therefore, the mixture weights have a stick-breaking form (see
Sethuraman, 1994\nocite{Sethuraman94}); later, we provide more details about
the lDP and these weights. In general, $\mathbf{x}$ can be a vector of any $%
p $ covariates, $\mathbf{x}=(x_{1},\ldots ,x_{p})$, and they respectively
correspond to (positive-valued)\ linear regression coefficients $\boldsymbol{%
\gamma }=(\gamma _{1},\ldots ,\gamma _{p})^{\intercal }$. For example, the
covariates may be dummy (0-1) test item indicators, describe examinee
characteristics (e.g., gender, race, and/or social economic status), and/or
describe other test characteristics (e.g., time at which item was
administered, item type, etc.).

As shown in equation (\ref{MixDist1}), the DDP-RM is based on an
infinite-mixture distribution $G_{\mathbf{x}}(\boldsymbol{\tau })$ for the
rating category thresholds $\boldsymbol{\tau }$. Therefore, conditionally on 
$\mathbf{x}$, the model can account for virtually all\ distributions of the
rating category thresholds ($\boldsymbol{\tau }_{h}$). These distributions
include unimodal distributions with small-variance, indicating an item is
free of DIF; unimodal distributions with larger-variance and/or skewness,
indicating an item with more uncertainty in rating category usage, and
possibly DIF; and multimodal distributions, which indicate the presence of
multiple latent clusters of examinees (i.e., DIF). Also, the shape and
location of the mixture distribution $G_{\mathbf{x}}$ can change flexibly as
a function of the covariates ($\mathbf{x}$). Therefore, at one extreme, the
mixture distribution $G_{\mathbf{x}}$ may be unimodal with small variance
for one value of the covariates $\mathbf{x}$, while for the other extreme,
the mixture distribution $G_{\mathbf{x}^{\prime }}$ may be highly skewed and
multimodal for a different value of the covariates $\mathbf{x}^{\prime }$.

The mixture distribution, $G_{\mathbf{x}}$, of our model is formed according
to our following novel modification of the local Dirichlet process (lDP)
(Chung \& Dunson, 2011\nocite{ChungDunson11}), which is described as
follows. First let 
\begin{equation*}
\mathcal{L}_{\mathbf{x}}=\left\{ h:d\left( \mathbf{x}^{\intercal }%
\boldsymbol{\gamma },h\right) \leq \psi (\mathbf{x})\right\} \subset
\{1,2,\ldots \}
\end{equation*}%
denote the subset of mixture component indices $h\in \mathbb{Z}^{+}$ having
fixed addresses $\{\Gamma _{h}\equiv h\}$ that are within a $\psi (\mathbf{x}%
)$-neighborhood around the linear predictor $\mathbf{x}^{\intercal }%
\boldsymbol{\gamma }$, $\pi _{l}\left( \mathbf{x}^{\intercal }\boldsymbol{%
\gamma }\right) $ is the $l^{th}$ ordered index in $\mathcal{L}_{\mathbf{x}}$%
, and $d\left( \mathbf{\cdot },\mathbf{\cdot }\right) $ is a chosen distance
measure (e.g., Euclidean). For example, if $\mathbf{x}^{\intercal }%
\boldsymbol{\gamma }=10$ and $\psi (\mathbf{x)}=2.5,$ then the covariate ($%
\mathbf{x}$)-dependent local subset becomes $\mathcal{L}_{\mathbf{x}%
}=\{8,9,10,11,12\},$ and $\pi _{1}\left( \mathbf{x}^{\intercal }\boldsymbol{%
\gamma }\right) =8,$ $\pi _{2}\left( \mathbf{x}^{\intercal }\boldsymbol{%
\gamma }\right) =9,...,\pi _{|\mathcal{L}_{\mathbf{x}}|}\left( \mathbf{x}%
^{\intercal }\boldsymbol{\gamma }\right) =12$, where $|\mathcal{L}_{\mathbf{x%
}}|$ is the cardinality of the set $\mathcal{L}_{\mathbf{x}}$. Under our
formulation of the lDP, the local variables are defined by $\boldsymbol{%
\upsilon }\left( \mathbf{x}^{\intercal }\boldsymbol{\gamma }\right) =\left\{
\upsilon _{h},h\in \mathcal{L}_{\mathbf{x}}\right\} $, in order to specify
the mixture weights in (\ref{MixDist1}) as having the covariate-dependent,
stick-breaking form%
\begin{equation}
\omega _{l}\left( \mathbf{x}^{\intercal }\boldsymbol{\gamma }\right)
=\upsilon _{\pi _{l}\left( \mathbf{x}^{\intercal }\boldsymbol{\gamma }%
\right) }\tprod\limits_{r<l}\left( 1-\upsilon _{\pi _{r}\left( \mathbf{x}%
^{\intercal }\boldsymbol{\gamma }\right) }\right) \text{, }
\label{lDP weights}
\end{equation}%
where the rating threshold atoms $\boldsymbol{\tau }\left( \mathbf{x}%
^{\intercal }\boldsymbol{\gamma }\right) =$ $\left\{ \tau _{h},h\in \mathcal{%
L}_{\mathbf{x}}\right\} $ are also covariate-dependent. We fix $\upsilon
_{\max (\mathcal{L}_{\mathbf{x}})}\left( \mathbf{x}^{\prime }\boldsymbol{%
\gamma }\right) \equiv 1$ to ensure that the mixture weights $\omega
_{l}\left( \mathbf{x}^{\intercal }\boldsymbol{\gamma }\right) $ sum to 1 for
each $\mathbf{x}$ (Chung \& Dunson, 2011\nocite{ChungDunson11}). In short,
our lDP\ forms stick-breaking mixture weights by selecting the strict subset
of stick-breaking parameters ($\left\{ \upsilon _{h}\right\} $) and atoms ($%
\left\{ \tau _{h}\right\} $) that are within the neighborhood centered
around (a linearized) $\mathbf{x}$. Then the mixture weights of Equation (%
\ref{lDP weights}) gives rise to a covariate-dependent mixing distribution
in equation (\ref{MixDist1}), which can be re-written as:%
\begin{equation}
G_{\mathbf{x}}(\cdot )=G_{\mathbf{x}}(\cdot ;\boldsymbol{\tau },\boldsymbol{%
\upsilon },\boldsymbol{\gamma },\boldsymbol{\psi })=\dsum\limits_{l=1}^{|%
\mathcal{L}_{\mathbf{x}}|}\omega _{l}\left( \mathbf{x}^{\intercal }%
\boldsymbol{\gamma }\right) \delta _{\boldsymbol{\tau }_{\pi _{l}\left( 
\mathbf{x}^{\intercal }\boldsymbol{\gamma }\right) }}\left( \cdot \right) 
\text{,}  \label{MixDist2}
\end{equation}%
where we denote $\boldsymbol{\tau }=(\boldsymbol{\tau }_{h})_{h=1}^{\infty }$%
, $\boldsymbol{\upsilon }=(\upsilon _{h})_{h=1}^{\infty }$, and $\boldsymbol{%
\psi }=(\psi (\mathbf{x}))_{\mathbf{x}\in \mathcal{X}}$. Based on this
specification, for two covariates $\mathbf{x}$ and $\mathbf{x}^{\prime }$,
the level of similarity between $\mathcal{L}_{\mathbf{x}}$ and $\mathcal{L}_{%
\mathbf{x}^{\prime }}$ determines the level of similarity between the two
corresponding mixing distribution $G_{\mathbf{x}}(\cdot )$ and $G_{\mathbf{x}%
^{\prime }}(\cdot )$, with the level of similarity controlled by the
parameters $(\boldsymbol{\gamma },\boldsymbol{\psi })$.

The DDP-RM is completed by the specification of the following prior
distributions:%
\begin{eqnarray*}
\theta _{t} &\sim &\text{\textrm{n}}\left( 0,\sigma ^{2}\right) ,\text{ }%
t=1,2,\ldots ,N; \\
\sigma ^{2} &\sim &\mathrm{ig}(\sigma ^{2}|a_{\sigma ^{2}},b_{\sigma ^{2}});
\\
\boldsymbol{\tau }_{h},\upsilon _{h} &\sim &\text{\textrm{n}}_{m_{j}}\left( 
\boldsymbol{\tau }|\mathbf{0},\mathbf{\Sigma }_{\tau }\right) \text{\textrm{%
beta}}\left( \upsilon |1,\alpha \right) ,\text{ }h=1,2,\ldots ; \\
\alpha ,\boldsymbol{\gamma } &\sim &\text{\textrm{ga}}\left( \alpha
|a_{\alpha },b_{\alpha }\right) \tprod\nolimits_{j=1}^{p}\mathrm{un}\left(
\gamma _{j}|a_{\gamma },b_{\gamma }\right) ; \\
\text{ \ }\psi (\mathbf{x}) &\sim &\text{\noindent \textrm{un}}\left(
a_{\psi },b_{\psi }\right) ,\text{ }\mathbf{x}\in \mathcal{X}.
\end{eqnarray*}%
If so desired, one may fix various model parameters to a particular constant
by making specific extreme choices of prior. For example, we can fix $\psi (%
\mathbf{x})$ to a constant $c$ by setting $a_{\psi }=b_{\psi }=c$ in the
uniform prior. Additionally, we can fix $\sigma ^{2}$ to 1, which is often
done in many IRT\ models, by taking $a_{\sigma ^{2}}\rightarrow \infty $ and 
$b_{\sigma ^{2}}$ $\rightarrow \infty $ in the inverse gamma prior.
Similarly, we can fix $\alpha $ to a fix value by appropriate choices of the
gamma parameters.

\subsection{Bayesian Posterior Inference of the DDP-RM}

For notational convenience, we denote a sample set of rating data by $%
\mathcal{D}_{n}=\{(\mathbf{x}_{i},y_{i})\}_{i=1}^{n=NJ}$, provided by $N$
examinees ($t=1,\ldots ,N$) on $J$\ test items ($j=1,\ldots ,J$), and with $%
n=NJ$ giving the total number of item responses in the data set. Each $%
y_{i}\in \mathcal{D}_{n}$ denotes a rating by a particular examinee on a
particular item. Additionally, as before, we denote the parameters of our
model by $\boldsymbol{\zeta }=(\boldsymbol{\theta },\sigma ^{2},\boldsymbol{%
\tau },\boldsymbol{\upsilon },\alpha ,\boldsymbol{\gamma },\boldsymbol{\psi }%
)$, with $\boldsymbol{\theta }=(\theta _{t})_{t=1}^{N}$, $\boldsymbol{\tau }%
=(\boldsymbol{\tau }_{h})_{h=1}^{\infty }$, $\boldsymbol{\upsilon }%
=(\upsilon _{h})_{h=1}^{\infty }$, $\boldsymbol{\gamma }=(\gamma
_{k})_{k=1}^{p}$, and $\boldsymbol{\psi }=(\psi (\mathbf{x}))_{\mathbf{x}\in 
\mathcal{X}}$.

According to standard arguments of probability theory involving Bayes'
theorem, given a data set $\mathcal{D}_{n}$ having likelihood $%
\tprod\nolimits_{i=1}^{n}f(y_{i}|\mathbf{x}_{i};\boldsymbol{\zeta })$ under
our model with parameters $\boldsymbol{\zeta }$, with a proper prior density 
$\pi (\boldsymbol{\zeta })$ defined over the space $\Omega _{\boldsymbol{%
\zeta }}$\ of $\boldsymbol{\zeta }$, the posterior density of $\boldsymbol{%
\zeta }$ is proper and is given by:%
\begin{equation*}
\pi (\boldsymbol{\zeta }|\mathcal{D}_{n})\propto
\dprod\nolimits_{i=1}^{n}P(y_{i}|\mathbf{x}_{i};\boldsymbol{\zeta })\pi (%
\boldsymbol{\zeta })
\end{equation*}%
up to a proportionality constant.\ Then the posterior predictive density of $%
Y$\ for a chosen $\mathbf{x}$ is given by:%
\begin{equation*}
f_{n}(y|\mathbf{x})=\int f(y|\mathbf{x};\boldsymbol{\zeta })\pi (\boldsymbol{%
\zeta }|\mathcal{D}_{n})\mathrm{d}\boldsymbol{\zeta },
\end{equation*}%
with this density corresponding to posterior predictive mean (expectation)\
and variance (Var)%
\begin{equation*}
\mathrm{E}_{n}(Y|\mathbf{x})=\tint yf_{n}(y|\mathbf{x})\mathrm{d}y,\text{ \
\ }\mathrm{Var}_{n}(Y|\mathbf{x})=\tint \{y-\mathrm{E}(Y|\mathbf{x}%
)\}^{2}f_{n}(y|\mathbf{x})\mathrm{d}y.
\end{equation*}%
Additionally, when investigating for DIF, it is of interest to infer
functionals of the posterior predictive mean $\mathrm{E}_{n}[G_{\mathbf{x}%
}(\cdot )]$ of the threshold mixture distribution $G_{\mathbf{x}}(%
\boldsymbol{\tau })$, such as its density. This posterior predictive mean is
defined by%
\begin{equation*}
\mathrm{E}_{n}[G_{\mathbf{x}}(\cdot )]=\int \int \int \int G_{\mathbf{x}%
}(\cdot ;\boldsymbol{\tau },\boldsymbol{\upsilon },\boldsymbol{\gamma },%
\boldsymbol{\psi })\pi (\boldsymbol{\tau },\boldsymbol{\upsilon },%
\boldsymbol{\gamma },\boldsymbol{\psi }|\mathcal{D}_{n})\mathrm{d}%
\boldsymbol{\tau }\mathrm{d}\boldsymbol{\upsilon }\mathrm{d}\boldsymbol{%
\gamma }\mathrm{d}\boldsymbol{\psi },
\end{equation*}%
given the marginal posterior density:%
\begin{equation*}
\pi (\boldsymbol{\tau },\boldsymbol{\upsilon },\boldsymbol{\gamma },%
\boldsymbol{\psi }|\mathcal{D}_{n})=\tint \tint \tint \pi \left( \boldsymbol{%
\zeta }|\mathcal{D}_{n}\right) \mathrm{d}\boldsymbol{\theta }\mathrm{d}%
\sigma ^{2}\mathrm{d}\alpha .
\end{equation*}%
In order to perform inference of functionals of the posterior density $\pi (%
\boldsymbol{\zeta }|\mathcal{D}_{n})$, including marginal posterior
densities, posterior predictive densities $f_{n}(y|\mathbf{x})$, the
posterior mean mixing distribution $\mathrm{E}_{n}[G_{\mathbf{x}}(\cdot )]$,
we make use of standard MCMC\ sampling methods for Bayesian infinite-mixture
models. These sampling methods are described by Kalli, Griffin, and Walker
(2011\nocite{kalli2011slice}). Appendix A provides more details about all
the conditional posterior distribution of the model, which are sampled at
each stage of the MCMC algorithm.

\subsection{Unique Features of the DDP-RM}

As mentioned, one unique feature of the DDP-RM is that it flexibly allows
the mixing distribution $G_{\mathbf{x}}$ to take on any shape, ranging from
unimodal with small variance, to highly multimodal with large variance.
Moreover, the mixing distribution $G_{\mathbf{x}}$ of the DDP-RM can
flexibly change as a function of the covariates $\mathbf{x}$. This
flexibility is enabled by a nonparametric specification of the mixing
distribution $G_{\mathbf{x}}$ according to a flexible infinite mixture
(involving an infinite number of parameters), with covariate-dependent
mixture weights (i.e., $\omega _{h}$) and thresholds (i.e., $\boldsymbol{%
\tau }_{h}$), as shown in equations (\ref{MixDist1}) and (\ref{MixDist2}).
In other words, the model makes no finite-parametric assumptions about this
mixing distribution, unlike traditional models, such as the assumption that
the mixing distribution is normally distributed and can be described by a
finite number of parameters (i.e., mean and variance). This assumption
implies the empirically-falsifiable assumption that the mixing distribution
is symmetric and unimodal. The DDP-RM, which is free from such limited
assumptions about the mixture distribution $G_{\mathbf{x}}$, allows for
accurate detection of rating scale category usage in the posterior
distribution of $G_{\mathbf{x}}(\cdot )$ for covariates $\mathbf{x}$ of
interest; for example, in the posterior means $\mathrm{E}_{n}[G_{\mathbf{x}%
}(\cdot )]$. This could help reveal when subsets or all category labels are
unclear, or when DIF\ is present.

Another unique feature of the DDP-RM is that it clusters item category
thresholds based on the similarity in the mixing distribution.\ This
similarity is captured through the neighborhood inducing parameter $%
\boldsymbol{\gamma }$. When two separate $\gamma $s have the same values,
the mixture components are the same for the covariates associates with the
two $\gamma $s. In the applications of our model for the simulated and real
data sets in Sections 3 and 4, we specify the covariates $\mathbf{x}$ by
dummy (0-1)\ test item indicators. Then, similar $\gamma $s would indicate
that the items associated with the $\gamma $s have similar mixing
distributions for the rating category thresholds.

\subsection{Model Assessment of Predictive Performance}

Given a set of data $\mathcal{D}_{n}$, one can use a a mean-squared
predictive error criterion, namely the $D(\underline{m})$ criterion (Gelfand
\& Ghosh, 1998\nocite{gelfand1998model}), to compare the predictive
performance among $\underline{M}$ different IRT\ rating models, with each
model indexed by $\underline{m}=1,...,\underline{M}.$ For a given model $%
\underline{m}\in \{1,...,\underline{M}\}$ under comparison, the criterion is
defined by: 
\begin{equation*}
D(\underline{m})=\dsum\limits_{i=1}^{n}\left[ y_{i}-\text{\textrm{E}}%
_{n}(Y_{i}|\text{$\mathbf{x}$}_{i},\underline{m})\right] ^{2}+\sum_{i=1}^{n}%
\text{\textrm{Var}}_{n}(Y_{i}|\text{$\mathbf{x}$}_{i},\underline{m})=\mathrm{%
GF}(\underline{m})+\mathrm{Pen}(\underline{m})
\end{equation*}%
In the right hand side of the equation above, the first term is a predictive
bias measure that indicates the goodness-of-fit ($\mathrm{GF}(\underline{m})$%
) of the model, to the sample data $\mathcal{D}_{n}$ at hand.\ The second
term is a penalty and is large when the model is either over-fitting or
under-fitting the data set $\mathcal{D}_{n}$. For all other comparison
models included in the present study, the E$_{n}(Y_{i}|\mathbf{x}$$_{i},%
\underline{m})$ and Var$_{n}(Y_{i}|\mathbf{x}$$_{i},\underline{m})$ are
derived from marginal maximum or conditional maximum likelihood parameter
estimates. For a non-Bayesian model having point estimate $\widehat{%
\boldsymbol{\zeta }}_{n}=\widehat{\boldsymbol{\zeta }}(\mathcal{D}_{n})$,
such as a maximum-likelihood estimate, the $D(\underline{m})$ criterion is
estimated via $\widehat{\mathrm{E}}_{n}(Y_{i}|\mathbf{x}_{i},\underline{m})=%
\mathrm{E}(Y_{i}|\mathbf{x}_{i},\underline{m},\widehat{\boldsymbol{\zeta }}%
_{n})$ and $\widehat{\mathrm{Var}}_{n}(Y_{i}|\mathbf{x}_{i},\underline{m})=%
\mathrm{Var}(Y_{i}|\mathbf{x}_{i},\underline{m},\widehat{\boldsymbol{\zeta }}%
_{n})$ ($i=1,\ldots ,n$) \ (Gelfand \& Ghosh, 1998\nocite{gelfand1998model}).

\section{Illustration of the DDP-RM on Simulated Data}

In this section, we provide a simulation study in order to demonstrate the
model's ability to correctly identify DIF due to the presence of multiple
latent examinee clusters, and to correctly identify the item free of DIF.

We generated item response data for 3000 examinees and 10 items, with each
item scored on a 0-2 rating scale, yielding a total of $n=30,000=3000\times
10$ rating observations. These data were generated according to the
parameters of a two-mixture Rasch logistic rating scale model, which are
described as follows. Each simulated\ examinee was assigned an ability $%
\theta $ parameter, according to an independent draw from a normal $\mathrm{n%
}(0,2.25)$ distribution. Additionally, each examinee was randomly assigned
to one of two clusters, with equal probability. As a result, 1505 and 1496
examinees were assigned to the first and second cluster, respectively.
Furthermore, each of the first nine items was specified as having no DIF in
the rating category thresholds, with the second threshold parameter $(\tau
_{2})$ being 1 unit larger than the first threshold parameter $(\tau _{1})$.
For example, the fifth item was assigned thresholds $\boldsymbol{\tau }%
=(\tau _{1}=-.5,\tau _{2}=.5)^{\intercal }$. Over all these nine items, the
category thresholds had range $(-2.3,2.3)$. In contrast, the tenth item was
specified to have DIF\ for the threshold parameter $\tau _{2}$, but no DIF
for the threshold $\tau _{1}$. Specifically, for this item, the first
threshold was specified as $\tau _{1}=-1.25$ for both examinee clusters. The
second threshold parameter was specified as $\tau _{2}=0$ for the first
examinee cluster, and specified as $\tau _{2}=2$ for the second examinee
cluster.

To analyze the simulated rating data using the DDP-RM, we made the following
model specifications for the purposes of demonstrating the model's ability
to differentiate between DIF and no-DIF items. First, we treated only two
items as having random (mixed)\ threshold parameters. They included the
fifth item, which was free of DIF, and the tenth item, which had DIF. For
each of the remaining eight items, the thresholds were treated as fixed
(non-mixed)\ parameters. Also, for the model, we specified covariates $%
\mathbf{x}$ as 0-1 dummy indicators of the 10 items. Thus we can write the
neighborhood size parameter as $\psi (\mathbf{x})=\psi _{j}$. Furthermore,
we assigned proper prior distributions to the model's parameters, namely $%
\theta _{t}\sim _{iid}$ \textrm{n}$\left( 0,\sigma ^{2}\right) $, $\sigma
^{2}\sim $ \textrm{ig}$\left( 1,1\right) $, $\boldsymbol{\tau }_{h}\sim
_{iid}$ \textrm{n}$\left( \mathbf{0},2\mathbf{I}_{m}\right) $, $\upsilon
_{h}\sim _{iid}$\textrm{beta}$\left( 1,\alpha \right) ,$ $\alpha \sim $ 
\textrm{ga}$\left( 1,1\right) $, $\gamma _{j}\sim _{iid}$\textrm{un}$\left(
1,745\right) $, while fixing $\psi _{j}=5$ for all items we treated as
random. For each of the eight items with fixed (non-mixed)\ threshold
parameters, the thresholds were assigned prior $\boldsymbol{\tau }\sim $ 
\textrm{n}$\left( \mathbf{0},10\mathbf{I}_{m}\right) $. We believe that
these prior distributions reflect priors that may be specified for typical
real-data applications of the DDP-RM, where little prior information is
available about the model parameters.

In order to perform Bayesian posterior estimation of the DDP-RM parameters,
we ran the MCMC\ sampling algorithm for 200,000 MCMC sampling iterations. We
discarded the first 100,000 MCMC samples (i.e., burn-in period), and saved
every fifth sample thereafter. This resulted in a total of 20,000 MCMC\
samples that we saved and used for posterior inference. We then used
standard procedures (Geyer, 2011\nocite{Geyer11}) to evaluate the
convergence of all MCMC\ samples to the posterior distribution of the model.
Univariate trace plots of the MCMC\ samples of model parameters showed good
mixing of the MCMC algorithm, in the sense that the MCMC\ samples of these
parameters seemed to stabilize and explore the support of the posterior
distribution with small auto-correlations. Also, we found that, for each
model parameter, the 20,000 saved MCMC\ samples led to a rather small 95\%\
Monte Carlo confidence interval (MCCI) for the parameter's marginal
posterior mean estimate, according to a consistent batch means estimator
(Jones, et al. 2006\nocite{jones2006fixed}). Over all model parameters, the
size of the 95\%\ MCCI half-width ranged between \TEXTsymbol{<}.01 and .02.
Hence, given all the results of the trace plots and 95\%\ MCCIs, we
generated a large-enough number of MCMC\ samples (200,000) to provide
reasonably-accurate posterior estimates of the model's parameters.

For the DDP-RM, the posterior mean estimates of the mixing distribution $G_{%
\mathbf{x}}(\boldsymbol{\tau })$, given covariates $\mathbf{x}$ (e.g. item
indicators), reveal how examinees used the rating categories. For the fifth
item, the top two panels of Figure 1 present the (marginal) posterior mean
density estimates of the mixture distributions $G_{\mathbf{x}}(\tau _{1})$
and $G_{\mathbf{x}}(\tau _{2})$, which correspond to the two rating
threshold parameters. As shown in the figure, for each of the two thresholds
of this fifth item, the marginal posterior mean density estimate was
unimodal with a very small variance. Thus, these estimates correctly shows
that the item has no DIF, in the sense that a single common set of category
thresholds applies to all examinees. That is, there is a single cluster of
examinees in terms of these thresholds. Moreover, the posterior mean
estimates of the thresholds were $\overline{\boldsymbol{\tau }}=(\overline{%
\tau }_{1}=-.44,\overline{\tau }_{2}=.43)^{\intercal }$, and are thus very
similar to the true data-generating values of $\boldsymbol{\tau }=(\tau
_{1}=-.5,\tau _{2}=.5)^{\intercal }$.

The bottom two panels of Figure 1 contain the estimated (marginal) posterior
densities of $G_{\mathbf{x}}(\tau _{1})$ and of $G_{\mathbf{x}}(\tau _{2})$
for the two rating threshold parameters associated with the tenth item. For
this item, the estimated marginal posterior density of the first threshold $%
G_{\mathbf{x}}(\tau _{1})$ is unimodal. Thus, this estimate correctly
indicates the presence of non-DIF for threshold parameter $\tau _{1}$. The
marginal posterior density estimate of the second threshold, however, is
bimodal. Hence, this estimate correctly indicates that there is DIF\ for
that item in that threshold. In other words, there are two latent clusters
(modes) of examinees in terms of that threshold parameter. Furthermore, the
first mode is slightly less than 0, and the second mode is approximately 2,
and are thus very close to the true modes (0 and 2, respectively)\ that were
used to simulate the rating data.

\section{Illustration of the DDP-RM on Real Data}

In this section, we illustrate the DDP-RM through the analysis of a real
data set obtained from the verbal aggression study (see De Boeck \& Wilson,
2004\nocite{de2004explanatory}), which was based on the Verbal Aggression
questionnaire. Moreover, we compare the predictive performance between the
DDP-RM and several other IRT\ rating models. This data set has been
frequently analyzed for the purposes of evaluating IRT\ models.
Specifically, this data set contains ratings of 24 items that were made by
each of 316 students (243 females and 73 males) who attended a
Dutch-speaking Belgian university. Each of the 24 items of the Verbal
Aggression questionnaire represents a type of verbal aggression (e.g.,
\textquotedblleft A bus fails to stop for me. I would want to
curse.\textquotedblright ), and can be categorized into a $2\times 2\times 3$
design: Behavior Mode (Want or Do) by Situation Type (Other-to-blame or
Self-to-blame) by Behavior Type (Curse, Scold, or Shout). Each item was
scored on a rating scale of 0 = no, 1 = perhaps, and 2 = yes.

\subsection{Model Specifications and MCMC\ Diagnostics}

To analyze the verbal aggression rating data using the DDP-RM, we treated
all items as having random (mixed)\ threshold parameters. Also, as before,
we specified the covariates $\mathbf{x}$ as 0-1 dummy indicators for the 24
Verbal Aggression items. Hence, we may rewrite the neighborhood size
parameter as $\psi (\mathbf{x})=\psi _{j}$. Furthermore, we assigned priors $%
\theta _{t}\sim _{iid}$ \textrm{n}$\left( 0,1\right) $, $\boldsymbol{\tau }%
_{h}\sim _{iid}$ \textrm{n}$\left( \mathbf{0},5\mathbf{I}_{m}\right) $, $%
\upsilon _{h}\sim _{iid}$\textrm{beta}$\left( 1,\alpha \right) ,$ $\alpha
\sim $ \textrm{ga}$\left( 1,1\right) $, $\gamma _{j}\sim _{iid}$ \textrm{un}$%
\left( 1,745\right) $, and $\psi _{j}\sim _{iid}\mathrm{un}\left(
.5,20\right) $, in our attempt to specify rather noninformative priors for
the model parameters. Finally, as is done with other IRT\ models, we assumed
that the item responses of the Verbal Aggression questionnaire are
independent, conditionally on all model parameters. Since each of the 24
questionnaire items can be classified according to $2\times 2\times 3$
design in terms of item type, there may be a concern that the data violate
this assumption. Though, if such a concern arises, then one can simply
specify additional covariates in the DDP-RM that describe the levels of this
design, so that it becomes more reasonable to assume conditional
independence under the (expanded)\ model. However, for the interests of
providing a simple illustration of the DDP-RM, we will analyze the data by
specifying the covariates $\mathbf{x}$ as 0-1 dummy indicators of the 24
questionnaire items.

To perform Bayesian posterior estimation of the DDP-RM parameters, we ran
the MCMC\ sampling algorithm for 200,000 MCMC sampling iterations. As
before, we discarded the first 100,000 MCMC samples (i.e., burn-in period),
and saved every fifth sample thereafter. This resulted in a total of 20,000
MCMC\ samples that we saved and used for posterior inference. Univariate
trace plots of the MCMC\ samples of model parameters showed good mixing of
the MCMC algorithm, in the sense that the MCMC\ samples of these parameters
seemed to stabilize and explore the support of the posterior distribution
with small auto-correlations. To provide more details, Figures 2 and 3
present the trace plots of the MCMC\ samples of the threshold parameters for
three items, and of the ability parameters for six examinees. Also, we found
that, for each model parameter, the 20,000 saved MCMC\ samples led to rather
small 95\%\ MCCIs for the marginal posterior mean estimates of various
parameters. For example, the size of the 95\% MC\ confidence interval
half-width had range $(.00,.03)$ for marginal posterior mean estimates of
examinee ability parameters, and had range $(.00,.03)$ for the marginal
posterior standard deviation of these parameters. Also, over all 24 item of
the Verbal Aggression questionnaire, the size of the 95\% MC\ confidence
interval half-width had range $(.02,.93)$ for the posterior mean and $.01$
to $.79$ for the posterior standard deviation for the neighborhood location $%
\gamma $ and for the for the neighborhood size $\psi $.

Over all model parameters, the size of the MCCI half-width typically ranged
between $.00$ and $.03$), with maximum value of $.05$. So given all the
results of the trace plots and 95\%\ MCCIs, it seems that we generated a
large-enough number of MCMC\ samples (200,000) to provide
reasonably-accurate posterior estimates of the model's parameters.

\subsection{Results}

Table \ref{Table Threshold summary} presents posterior mean and standard
deviation estimates of the category threshold parameters for each of the 24
items. As shown, the posterior means ranged from $-0.68$ to $3.32$. Similar
to conclusions by others (e.g., De Boeck \& Wilson, 2004\nocite%
{de2004explanatory}), Item 21 was found to be the most difficult to endorse,
as it attained the the largest posterior means for the category thresholds.\
Item 4 was the easiest to endorse, as it had the smallest posterior means
for the category thresholds.

For three of the Verbal Aggression questionnaire items, Figure 4 contains
the marginal posterior mean density estimates of $G_{\mathbf{x}}(\tau _{1})$
and $G_{\mathbf{x}}(\tau _{2})$ for the two rating threshold parameters. As
shown, Items 1 and 23 exhibit greater variability in their rating category
thresholds compared to Item 2. For Item 1, the marginal posterior mean
density estimate of the first threshold ($\tau _{1}$)\ and the second
threshold ($\tau _{2}$) is tri-modal and bimodal, respectively. Thus, the
item contains DIF, in the sense that there are three distinct latent
clusters of examinees with respect to threshold $\tau _{1}$, and two
distinct latent clusters of examinees with respect to threshold $\tau _{2}$.
For Item 23, the marginal posterior mean density estimate is bimodal for
threshold parameter $\tau _{1}$ and for threshold parameter $\tau _{2}$.
Hence, this item also contains DIF. On the other hand, for Item 2, the
marginal posterior mean density estimate for each threshold is unimodal with
small variance. Thus, these estimates suggest no DIF, and indicate that
there is a single cluster of examines in terms of these threshold parameters.

For all 24 items of the Verbal Aggression questionnaire, Table \ref{Table
Threshold summary} presents the marginal posterior mean, standard deviation,
and modes of the threshold distributions $G_{\mathbf{x}}(\tau _{1})$ and $G_{%
\mathbf{x}}(\tau _{2})$. As shown, 21 of the 24 items are unimodal. The
multimodal items, such as Items 21 and Item 23, may be referred to content
experts on verbal aggression, so that they can provide further explanation
as to why they are exhibiting DIF, and provide advice as to how to modify
and improve the questionnaire for its future use. However, in the case the
one must retain all possible data, such items do not pose problems for the
DDP-RM itself because the model accounts for DIF, and therefore produces
posterior parameter estimates (e.g., of examinee ability parameters) after
controlling for any DIF. In contrast, for an IRT\ model that assumes no DIF
in the rating threshold parameters, the presence of DIF\ in the data will
lead to misleading parameter estimates. As mentioned in the Introduction
section, such estimates would wrongly indicate that a single set of rating
category threshold estimates applies for all examinees, for each test item.
In turn, this may lead to misleading examinee ability estimates.

The DDP-RM also provides information about the similarities in mixing
distributions over the 24 questionnaire items, through the neighborhood
location and size parameters (i.e., $\gamma _{j}$ and $\psi _{j}$,
respectively). Over all the 24 items, the marginal posterior mean estimates
of the neighborhood location parameter $\gamma _{j}$ ranged from $6.0$ to $%
255.6$, whereas the marginal posterior mean estimates of neighborhood size
parameter $\psi _{j}$ ranged from $7.5$ to $19.8$. In terms of the posterior
posterior means, the items had noticeably different neighborhood locations
and sizes, indicating that the items differed in terms of the mixing
distribution $G_{\mathbf{x}}(\boldsymbol{\tau })$. The box-plots in Figure 5
presents the marginal posterior median and interquartile range estimates for
the neighborhood location and neighborhood size parameters, for each of the
24 questionnaire items.

Finally, over the 316 examinees (students), the marginal posterior mean
estimate of the ability parameter $\theta _{t}$ had range $(-2.37,3.74)$,
along with a mean $-0.02$ and standard deviation $1.01$.

\subsection{Model Comparisons}

In this section, for the Verbal Aggression data set, we compared the
predictive performance between the DDP-RM, and other well-known IRT\ rating
models. The other models include the partial credit model (PCM)\ (Masters,
1982\nocite{masters1982rasch}), the generalized partial credit model (GPCM)\
(Muraki, 1992\nocite{Muraki92}), the rating scale model (RSM)\ (Andrich, 1978%
\nocite{andrich1978rating}), the graded response model (GRM)\ (Samejima, 1969%
\nocite{Samejima69}), the nominal response model (NRM)\ (Bock, 1972), the
mixture partial credit model (mix-PCM)\ (Rost, 1991\nocite{rost1991logistic}%
), and a covariate-independent DP mixture PCM\ model that treated the
category thresholds as random. All models except the latter two were fit
using IRTPRO 2.1 (Cai, Thissen, \& du Toit, 2011\nocite{cai2011irtpro}). The
mix-PCM was fit in WINMIRA 2001 (von Davier, 2001\nocite{von2001winmira}).
Among the one-, two-, three-, four-, and five-mixture PCMs, the 3-mixture
PCM\ displayed the best predictive performance, according to the Akaike
Information Criterion (Akaike, 1973\nocite{Akaike73}), an index of model
predictive fit. Thus, we report the predictive performance of the
three-mixture PCM. The DP mixture PCM\ model was fit using code we wrote in
MATLAB (2012, The MathWorks, Natick, MA). For this model, the baseline
distribution for the set of $m$ thresholds was distributed as a multivariate
normal distribution with density function \textrm{n}$\left( \boldsymbol{\tau 
}|\mathbf{0},\mathbf{I}_{m}\right) $, the examinee abilities were assigned a
normal \textrm{n}$\left( 0,1\right) $ prior distribution, and the precision
parameter $\alpha $ was fixed to 1. For the DDP-RM and the DP mixture PCM,
we estimated the posterior distribution of parameters using 200,000 MCMC\
sampling iterations, as before. In each of these cases, the size of the
95\%\ MCCI\ half width was generally less than 1, over all model parameters.

Table \ref{Model Compare}\ presents the $D(\underline{m})$ mean-squared
error predictive criterion for each of the IRT\ models used to analyze the
Verbal Aggression data set. As shown, the DDP-RM outperformed all comparison
models, by at least 49 $D(\underline{m})$ units. Moreover, in terms of the $%
D(\underline{m})$, there was no overlap between any two of the models after
accounting for the 95\%\ MCCI of the $D(\underline{m})$ estimate. In all,
the three mixture models outperformed the traditional, non-mixture models.
This result suggests that more than one latent class is present in the data
set. Nevertheless, the finite-mixture Rasch PCM model, while outperforming
the traditional models, was still bested by the two infinite-mixture models.
The DDP-RM outperforming the DP-mixture PCM suggests that all items do not
share a common mixing distribution. As mentioned in the previous subsection,
the items have noticeably different posterior mean estimates for the
neighborhood location item parameters, and for the neighborhood size item
parameters.

\section{Conclusions}

We have introduced a novel Bayesian nonparametric rating scale IRT\ model
named the DDP-RM. This is an infinite-mixture model that is based on the
local Dirichlet process formulation of the DDP. The model, through posterior
mean estimates of the mixing distribution for the threshold parameters,
describes how the examinees used the rating categories. Specifically, the
posterior number of modes in the mixing distribution reveals the number of
clusters (groups)\ of examinees in terms of an item category thresholds.
Moreover, using a real data set that is well-known in the psychometric
field, we demonstrated that the new model provides a substantially-better
predictive fit of the rating data compared to other IRT\ models commonly
used.

In future research, the DDP-RM can be extended by assigning a nonparametric
prior for the ability distribution, such as a DP prior (San Martin, Jara,
Rolin, \& Mouchart, 2011\nocite{san2011bayesian}). Also, it would be of
interest to extend the model by specifying $G_{\mathbf{x}}\left( \boldsymbol{%
\tau }\right) $ as a more flexible, infinite mixture. For example,
Karabatsos and Walker (2012\nocite{KarabatsosWalker12c}) proposed novel
mixture weights that are based on an infinite-ordered probits regression
model, with covariate dependence in the mean and in the variance of the
probits. Alternatively, the infinite number of\ mixture weights can be
specified by a covariate-dependent version of normalized random measures
(Regazzini, Lijoi \&\ Pr\"{u}nster, 2003\nocite{RegazziniLijoiPrunster03};
Lijoi, Me\~{n}a, \& Pr\"{u}nster, 2005, 2007\nocite{LijoiMenaPrunster05}%
\nocite{LijoiMenaPrunster07}; James, Lijoi, \&\ Pr\"{u}nster, 2009\nocite%
{JamesLijoiPrunster09}).

\bigskip

\bigskip \noindent {\LARGE References}

\bigskip

\begin{description}
\item Akaike, H. (1973). Information theory and the an extension of the
maximum likelihood principle. In B. Petrov \& F. Csaki (Eds.), \textit{%
Second international symposium on information theory} (pp. 267-281).
Budapest: Academiai Kiado.

\item Andrich, D. (1978). A rating formulation for ordered response
categories. \textit{Psychometrika}, \textit{43}, 561-573.

\item Cai, L., Toit, S. du, \& Thissen, D. (2011). \textit{IRTPRO: Flexible,
multidimensional, multiple categorical IRT modeling}. Chicago, IL:
Scientific Software International.

\item Chung, Y., \& Dunson, D. B. (2011). The local Dirichlet process. 
\textit{Annals of the Institute of Statistical Mathematics}, \textit{63},
59-80.

\item De Boeck, P., \& Wilson, M. (2004). \textit{Explanatory item response
models: A generalized linear and nonlinear approach}. Springer.

\item Duncan, K. A., \& MacEachern, S. N. (2008). Nonparametric Bayesian
modelling for item response. \textit{Statistical Modelling}, \textit{8},
41-66.

\item Escobar, M. D., \& West, M. (1995). Bayesian density estimation and
inference using mixtures. \textit{Journal of the American Statistical
Association}, \textit{90}, 577-588.

\item Frick, H., Strobl, C., Leisch, F., \& Zeileis, A. (2012). Flexible
Rasch mixture models with package psychomix. \textit{Journal of Statistical
Software}, \textit{48}, 1-25.

\item Gelfand, A. E., \& Ghosh, S. K. (1998). Model choice: A minimum
posterior predictive loss approach. \textit{Biometrika}, \textit{85}, 1-11.

\item Geyer, C. (2011). Introduction to MCMC. In S. Brooks, A. Gelman, G.
Jones, \& X. Meng (Eds.), \textit{Handbook of Markov Chain Monte Carlo} (pp.
3-48). Boca Raton, FL: CRC.

\item James, L. F., Lijoi, A., \& Pr\"{u}nster, I. (2009). Posterior
analysis for normalized random measures with independent increments. \textit{%
Scandinavian Journal of Statistics}, \textit{36}, 76-97.

\item Jones, G. L., Haran, M., Caffo, B. S., \& Neath, R. (2006).
Fixed-width output analysis for Markov chain Monte Carlo. \textit{Journal of
the American Statistical Association}, \textit{101}, 1537-1547.

\item Kalli, M., Griffin, J. E., \& Walker, S. G. (2011). Slice sampling
mixture models. \textit{Statistics and Computing}, \textit{21}, 93-105.

\item Karabatsos, G., \& Walker, S. G.(2012). Adaptive-modal Bayesian
nonparametric regression. \textit{Electronic Journal of Statistics}, \textit{%
6}, 2038-2068.

\item Karabatsos, G., \& Walker, S. G. (2013 to appear). Bayesian
nonparametric IRT. In W. van der Linden \& R. Hambleton (Eds.), \textit{%
Handbook of item response theory: Models, statistical tools, and applications%
}. New York: Taylor and Francis.

\item Lijoi, A., Mena, R. H., \& Pr\"{u}nster, I.(2005). Hierarchical
mixture modeling with normalized inverse-Gaussian priors. \textit{Journal of
the American Statistical Association}, \textit{100}, 1278-1291.

\item Lijoi, A., Mena, R. H., \& Pr\"{u}nster, I. (2007). Controlling the
reinforcement in Bayesian nonparametric mixture models. \textit{Journal of
the Royal Statistical Society}, \textit{Series B}, \textit{69}, 715-740.

\item Lord, F. M. (1980). \textit{Application of item response theory to
practical testing problems}. Lawrence Erlbaum.

\item MacEachern, S. N.(1999). Dependent nonparametric processes. \textit{%
Proceedings of the Bayesian Statistical Sciences Section of the American
Statistical Association}, 50-55.

\item MacEachern, S. N. (2000). \textit{Dependent Dirichlet Processes}
(Tech. Rep.). The Ohio State University: Department of Statistics.

\item MacEachern, S. N. (2001). Decision theoretic aspects of dependent
nonparametric processes. In E. George (Ed.), \textit{Bayesian methods with
applications to science, policy and official statistics} (pp. 551-560).
Creta: International Society for Bayesian Analysis.

\item Masters, G. N. (1982). A Rasch model for partial credit scoring. 
\textit{Psychometrika}, \textit{47}, 149-174.

\item McLachlan, G., \& Peel, D. (2000). \textit{Finite mixture models}.
Wiley-Interscience.

\item Miyazaki, K., \& Hoshino, T. (2009). A Bayesian semiparametric item
response model with Dirichlet process priors. \textit{Psychometrika}, 
\textit{74}, 375-393.

\item Muraki, E. (1992). A generalized partial credit model: Application of
an EM algorithm. \textit{Applied Psychological Measurement},\textit{\ 16},
159-176.

\item Regazzini, E., Lijoi, A., \& Pr\"{u}nster, I.(2003). Distributional
results for means of normalized random measures with independent increments. 
\textit{Annals of Statistics}, \textit{31}, 560-585.

\item Roberts, G. O., \& Rosenthal, J. S. (2009). Examples of adaptive MCMC. 
\textit{Journal of Computational and Graphical Statistics}, \textit{18},
349-367.

\item Rost, J. (1991). A logistic mixture distribution model for polytomous
item responses. \textit{Journal of Mathematical and Statistical Psychology}, 
\textit{44}, 75-92.

\item Samejima, F. (1969). Estimation of latent ability using a response
pattern of graded scores. \textit{Psychometrika Monograph Supplement}.

\item Samejima, F. (1972). A general model for free response data.\textit{\
Psychometrika Monograph}, \textit{18}.

\item San Mart\'{\i}n, E., Jara, A., Rolin, J. M., \& Mouchart, M. (2011).
On the Bayesian nonparametric generalization of IRT-type models. \textit{%
Psychometrika}, \textit{76}, 385-409.

\item Seol, H. (1999). Detecting differential item functioning with five
standardized item-fit indices in the Rasch model. \textit{Journal of Outcome
Measurement}, \textit{3}, 233-247.

\item Sethuraman, J. (1994). A constructive definition of Dirichlet priors. 
\textit{Statistica Sinica}, \textit{4}, 639-650.

\item Smit, A., Kelderman, H., \& Flier, H. van der. (2003). Latent trait
latent class analysis of an Eysenck Personality Questionnaire. \textit{%
Methods of Psychological Research Online}, \textit{8}, 23-50.

\item Smith, R. M., \& Suh, K. K. (2003). Rasch fit statistics as a test of
the invariance of item parameter estimates. \textit{Journal of Applied
Measurement}, \textit{4}, 153-163.

\item von Davier, M. (2001). \textit{WINMIRA 2001}. [Computer software]. St.
Paul, MN: Assessment Systems Corporation.

\item von Davier, M., \& Yamamoto, K. (2004). Partially observed mixtures of
IRT models: An extension of the generalized partial-credit model. \textit{%
Applied Psychological Measurement}, \textit{28}, 389-406.

\item Wright, B., \& Masters, G. (1982). \textit{Rating scale analysis}.
Chicago: MESA Press.
\end{description}

\newpage

\begin{center}
\appendix\textbf{APPENDIX A: MCMC Sampling Methods}
\end{center}

We implement the MCMC\ sampling method of Kalli et al., (2011\nocite%
{kalli2011slice}) to estimate our infinite-mixture IRT\ model. This MCMC
sampling method involves introducing strategic latent variables in order to
implement exact MCMC algorithms for the estimation of the model's posterior
distribution. That is, for our DDP-RM\ (Section \ref{Section on DDP-RM}), we
introduce the latent variables ($u_{i},$ $z_{i}\in \mathbb{Z}$)$_{i=1}^{n}$
and a decreasing function $\xi _{h}=\exp (-h),$ so that the model's data
likelihood can be written as the joint distribution:%
\begin{equation}
\dprod\limits_{i=1}^{n}f(u_{i},z_{i},y_{i}|\mathbf{x};\boldsymbol{\zeta }%
)=\dprod\limits_{i=1}^{n}\left\{ \mathbb{I}\left( 0<u_{i}<\xi
_{z_{i}}\right) \xi _{z_{i}}^{-1}f\left( y_{i}|\theta _{t(i)},\boldsymbol{%
\tau }_{z_{i}}\right) \omega _{z_{i}}\left( \mathbf{x}^{\prime }\boldsymbol{%
\gamma }\right) \right\} ,  \label{Slice Sampling Likelihood}
\end{equation}%
where $\theta _{t(i)}$ denotes the ability of examinee $t$ who provided the
rating $y_{i}$, and where $\mathbb{I}\left( \cdot \right) $ is the indicator
function. Marginalizing over each of the latent variables $(u_{i},z_{i})$ in
Equation \ref{Slice Sampling Likelihood}, for each $i=1,...n$, returns the
original likelihood,%
\begin{equation*}
\dprod\limits_{i=1}^{n}\left\{ \dsum\limits_{h=1}^{\infty }f\left( y|\theta
_{t(i)}\mathbf{,}\boldsymbol{\tau }_{h}\right) \omega _{h}\left( \mathbf{x}%
^{\intercal }\boldsymbol{\gamma }\right) ,\right\} ,
\end{equation*}%
of our infinite-dimensional IRT\ model. Thus, provided the latent variables,
the model can be characterized as a finite-dimensional model, which in turn,
permits the use of standard MCMC\ methods to sample the model's full joint
posterior distribution. Given all variables, save the $(z_{i})_{i=1}^{n}$,
the choice of each $z_{i}$ has minimum 1 and maximum $N_{\max }$, where $%
N_{\max }=\max_{i}\left[ \max_{h}\mathbb{I}\left( u_{i}<\xi _{h}\right) h%
\right] $.

Specifically, for each $i=1,...,n$ and $t=1,...,T$, each of the model
parameters is sampled from its corresponding full conditional posterior
distribution at each stage $s$ ($s=1,...,S$) of the MCMC algorithm. We
assume the prior form as in the empirical illustration of our model, in the
analysis of the verbal aggression data set, as in Section 4. \noindent The
full conditional posterior distribution for each block of model parameters
are as follows:

\begin{enumerate}
\item $\pi \left( u_{i}|...\right) =$ \textrm{un}$\left( u_{i}|0,\xi
_{z_{i}}\right) ;$

\item $\pi \left( z_{i}=h|...\right) \propto \mathbb{I}\left( u_{i}<\xi
_{h}\right) \xi _{h}^{-1}f\left( y_{i}|\theta _{t(i)},\boldsymbol{\tau }%
_{h}\right) \omega _{h}\left( \mathbf{x}^{\prime }\boldsymbol{\gamma }%
\right) ,$ $h=1,...,N_{\max };$

\item $\pi \left( \theta _{t}|...\right) \propto $ \textrm{n}$(\theta
|0,\sigma ^{2})\tprod\nolimits_{i:t(i)=t}f\left( y_{i}|\theta _{t(i)},%
\boldsymbol{\tau }_{z_{i}}\right) ;$

\item $\pi \left( \sigma ^{2}|...\right) =$ \textrm{ig}$(\sigma
^{2}|a_{\sigma ^{2}}+N/2,b_{\sigma ^{2}}+\frac{1}{2}\tsum\nolimits_{i=1}^{N}%
\theta _{t}^{2});$

\item $\pi \left( \boldsymbol{\gamma }|...\right) \propto $ $%
\{\tprod\nolimits_{j=1}^{p}$\textrm{un}$\left( \gamma _{j}|a_{\gamma
},b_{\gamma }\right) \}\tprod\nolimits_{i=1}^{n}\upsilon
_{z_{i}}\tprod\nolimits_{l=1}^{z_{i}-1}\left( 1-\upsilon _{l}\right) ;$

\item $\pi \left( \psi (\mathbf{x})|...\right) \propto $ \textrm{un}$(\psi (%
\mathbf{x})|a_{\psi },b_{\psi })\tprod\nolimits_{i=1}^{n}\upsilon
_{z_{i}}\tprod\nolimits_{l=1}^{z_{i}-1}\left( 1-\upsilon _{l}\right) ;$

\item $\pi \left( \boldsymbol{\tau }_{h}|...\right) \propto $ \textrm{n}$%
_{m}\left( \boldsymbol{\tau }_{h}|\mathbf{0},\mathbf{\Sigma }_{\tau }\right)
\tprod\nolimits_{i\in h}f\left( y_{i}|\theta _{t\left( i\right) },%
\boldsymbol{\tau }_{z_{i}}\right) ,h=1,...,N_{\max };$

\item $\pi \left( \upsilon _{h}|...\right) =$ \textrm{beta}$\left( \upsilon
_{h}\left\vert 1+\tsum\limits_{i=1}^{n}\mathbb{I(}z_{i}=h\text{ }\&\text{ }%
z_{i}\neq \max \{\mathcal{L}_{\mathbf{x}}\}),\alpha +\tsum\limits_{i=1}^{n}%
\mathbb{I}\left( z_{i}>h\right) \right. \right) ,$ $h=1,...,N_{\max };$

\item $\pi \left( \alpha |...\right) =$ \textrm{ga}$(\alpha |a_{\alpha
}+n_{clus}-\mathbb{I}(u>\{O/(1+O)\}),\{b_{\alpha }-\log (\eta )\}^{-1})$,
given draws $\eta \sim $ \textrm{beta}$(\alpha +1,n)$, $u\sim $ \textrm{un}$%
(0,1)$ , and $O=(a_{\alpha }+n_{clus}-1)/(\{b_{\alpha }-\log (\eta )\}n)$,
where $n_{clus}$ is the number of unique $z_{i}$, over ($i=1,\ldots ,n$)
(Escobar \& West, 1995, p.584\nocite{EscobarWest95}).
\end{enumerate}

Standard MCMC Gibbs sampling methods can be used to sample the full
conditionals in Steps 1, 2, 4, 8, and 9. The full conditionals in Steps 3,
5, 6, and 7 are each sampled using an adaptive random-walk
Metropolis-Hastings algorithm (Roberts \&\ Rosenthal, 2009\nocite%
{RobertsRosenthal09}). The above 9-step sampling algorithm is repeated a
large number $S$ of times to construct a discrete-time Harris ergodic Markov
chain $\left\{ \boldsymbol{\zeta }^{(s)}=(\boldsymbol{\theta },\sigma ^{2},%
\boldsymbol{\tau },\boldsymbol{\upsilon },\alpha ,\boldsymbol{\gamma },%
\boldsymbol{\psi })^{(s)}\right\} _{s=1}^{S}$, having a posterior
distribution $\Pi \left( \boldsymbol{\zeta }|\mathcal{D}_{n}\right) $ as its
stationary distribution, provided that a proper prior is assigned to $%
\boldsymbol{\zeta }\mathbf{.}$

The posterior predictive density $f_{n}(y|\mathbf{x})$, and the posterior
mean of the mixing distribution $\mathrm{E}_{n}[G_{\mathbf{x}}(\cdot )]$,
and the functionals thereof (e.g., a kernel density estimate), can each be
estimated as simple by-products of the MCMC\ algorithm. In order to estimate
the posterior predictive density $f_{n}(y|\mathbf{x})$, a step is added to
the MCMC\ algorithm, to sample from the full conditional posterior
distribution $f\left( y_{i}|\theta _{t(i)},\boldsymbol{\tau }_{z_{i}}\right) 
$, which is a multinomial distribution defined by the Rasch partial credit
model. A MCMC\ sample of $\mathrm{E}_{n}[G_{\mathbf{x}_{i}}(\cdot )]$ is
given by $\boldsymbol{\tau }_{z_{i}}$.

We have written MATLAB (2012, The MathWorks, Natick, MA) code that
implements the MCMC sampling algorithm. The analysis of the verbal
aggression data took approximately 24 hours, using a Dell Precision T3600,
3.2 GHz 6-core, and 32 gigs of RAM.\newpage

\QTP{Body Math}
\begingroup%

\renewcommand{\baselinestretch}{1}
%

\QTP{Body Math}
\begin{table}[H] \centering%
\begin{tabular}{lccccll}
\hline
& \multicolumn{2}{c}{$\tau _{1}$} & \multicolumn{2}{c}{$\tau _{2}$} &  &  \\ 
\cline{2-5}
Item & Mean & SD & Mean & SD & Modes $\tau _{1}$ & Modes $\tau _{2}$ \\ 
\cline{2-7}
1: bus-want-curse & \multicolumn{1}{l}{$-.42$} & \multicolumn{1}{l}{$1.27$}
& $-.03$ & \multicolumn{1}{l}{$1.87$} & $-.1,-.9,1.4$ & $.6,-.5$ \\ 
2: bus-want-scold & \multicolumn{1}{l}{$.06$} & \multicolumn{1}{l}{$.83$} & $%
.20$ & \multicolumn{1}{l}{$.85$} & $.1$ & $.2$ \\ 
3: bus-want-shout & \multicolumn{1}{l}{$.28$} & \multicolumn{1}{l}{$.85$} & $%
1.09$ & \multicolumn{1}{l}{$1.00$} & $.4$ & $1.4$ \\ 
4: train-want-curse & \multicolumn{1}{l}{$-.68$} & \multicolumn{1}{l}{$1.47$}
& $.09$ & \multicolumn{1}{l}{$1.55$} & $-.3,-.9$ & $.6,-1.7,$ \\ 
& \multicolumn{1}{l}{} & \multicolumn{1}{l}{} &  & \multicolumn{1}{l}{} &  & 
$-2.3,-.5,1.9$ \\ 
5: train-want-scold & \multicolumn{1}{l}{$-.10$} & \multicolumn{1}{l}{$.25$}
& $.25$ & \multicolumn{1}{l}{$.26$} & $-.2$ & $.2$ \\ 
6: train-want-shout & \multicolumn{1}{l}{$.33$} & \multicolumn{1}{l}{$1.74$}
& $.67$ & \multicolumn{1}{l}{$1.21$} & $-.2$ & $.6$ \\ 
7: grocery-want-curse & \multicolumn{1}{l}{$-.14$} & \multicolumn{1}{l}{$.87$%
} & $1.11$ & \multicolumn{1}{l}{$1.43$} & $-.4$ & $1.5$ \\ 
8: grocery-want-scold & \multicolumn{1}{l}{$.82$} & \multicolumn{1}{l}{$.29$}
& $2.01$ & \multicolumn{1}{l}{$.42$} & $.8$ & $2.0$ \\ 
9: grocery-want-shout & \multicolumn{1}{l}{$1.52$} & \multicolumn{1}{l}{$.52$%
} & $2.75$ & \multicolumn{1}{l}{$.70$} & $1.6$ & $2.8,3.8$ \\ 
10: operator-want-curse & \multicolumn{1}{l}{$-.63$} & \multicolumn{1}{l}{$%
.52$} & $.70$ & \multicolumn{1}{l}{$.56$} & $-.8$ & $.7$ \\ 
11: operator-want-scold & \multicolumn{1}{l}{$.63$} & \multicolumn{1}{l}{$%
.47 $} & $1.29$ & \multicolumn{1}{l}{$.59$} & $.7$ & $1.4$ \\ 
12: operator-want-shout & \multicolumn{1}{l}{$1.28$} & \multicolumn{1}{l}{$%
1.05$} & $1.70$ & \multicolumn{1}{l}{$1.16$} & $1.6$ & $2.0$ \\ 
13: bus-do-curse & \multicolumn{1}{l}{$-.61$} & \multicolumn{1}{l}{$.46$} & $%
.21$ & \multicolumn{1}{l}{$.47$} & $-.6$ & $.2$ \\ 
14: bus-do-scold & \multicolumn{1}{l}{$.14$} & \multicolumn{1}{l}{$.72$} & $%
.63$ & \multicolumn{1}{l}{$1.2$} & $-.06$ & $.84$ \\ 
15: bus-do-shout & \multicolumn{1}{l}{$1.15$} & \multicolumn{1}{l}{$.86$} & $%
1.69$ & \multicolumn{1}{l}{$1.58$} & $1.38,.22$ & $2.23$ \\ 
16: train-do-curse & \multicolumn{1}{l}{$-.25$} & \multicolumn{1}{l}{$.92$}
& $.20$ & \multicolumn{1}{l}{$1.24$} & $-.46$ & $.33$ \\ 
17: train-do-scold & \multicolumn{1}{l}{$.48$} & \multicolumn{1}{l}{$.77$} & 
$1.04$ & \multicolumn{1}{l}{$1.35$} & $.46$ & $1.29$ \\ 
18: train-do-shout & \multicolumn{1}{l}{$1.62$} & \multicolumn{1}{l}{$1.00$}
& $2.17$ & \multicolumn{1}{l}{$1.2$} & $1.94$ & $2.47$ \\ 
19: grocery-do-curse & \multicolumn{1}{l}{$.89$} & \multicolumn{1}{l}{$.64$}
& $2.12$ & \multicolumn{1}{l}{$.93$} & $1.02$ & $2.25$ \\ 
20: grocery-do-scold & \multicolumn{1}{l}{$.96$} & \multicolumn{1}{l}{$.38$}
& $2.24$ & \multicolumn{1}{l}{$.56$} & $1.10$ & $2.21$ \\ 
21: grocery-do-shout & \multicolumn{1}{l}{$2.87$} & \multicolumn{1}{l}{$.52$}
& $3.31$ & \multicolumn{1}{l}{$.77$} & $2.92$ & $3.22$ \\ 
22: operator-do-curse & \multicolumn{1}{l}{$-.22$} & \multicolumn{1}{l}{$.86$%
} & $.80$ & \multicolumn{1}{l}{$1.34$} & $-.48$ & $1.06$ \\ 
23: operator-do-scold & \multicolumn{1}{l}{$.61$} & \multicolumn{1}{l}{$1.83$%
} & $1.01$ & \multicolumn{1}{l}{$1.27$} & $-.15,$ $2.67$ & $.64,1.05$ \\ 
24: operator-do-shout & \multicolumn{1}{l}{$2.06$} & \multicolumn{1}{l}{$.07$%
} & $2.56$ & \multicolumn{1}{l}{$.89$} & $2.17$ & $2.45$ \\ \hline
\end{tabular}%
\caption{For the DDP-RM, the posterior estimates of the ordered category threshold parameters, by item. 
For the posterior mean and SD estimates, the 95 percent MCCI half-width typically ranged between .00 to 03, with maximum .05.}%
\label{Table Threshold summary}%
\end{table}%
\bigskip

\QTP{Body Math}
\newpage

\begin{table}[H] \centering%
\begin{tabular}{llll}
\hline
Model ($\underline{m}$) & $D(\underline{m})$ & GF$(\underline{m})$ & Pen$(%
\underline{m})$ \\ \hline
DDP-RM & \multicolumn{1}{c}{$4984$} & \multicolumn{1}{c}{$2008$} & 
\multicolumn{1}{c}{$2976$} \\ 
DP-PCM & \multicolumn{1}{c}{$5033$} & \multicolumn{1}{c}{$2077$} & 
\multicolumn{1}{c}{$2956$} \\ 
3-Mixture PCM & \multicolumn{1}{c}{$5163$} & \multicolumn{1}{c}{$2485$} & 
\multicolumn{1}{c}{$2679$} \\ 
PCM & \multicolumn{1}{c}{$5716$} & \multicolumn{1}{c}{$2783$} & 
\multicolumn{1}{c}{$2934$} \\ 
GPCM & \multicolumn{1}{c}{$5686$} & \multicolumn{1}{c}{$2774$} & 
\multicolumn{1}{c}{$2912$} \\ 
RSM & \multicolumn{1}{c}{$5726$} & \multicolumn{1}{c}{$2791$} & 
\multicolumn{1}{c}{$2936$} \\ 
NRM & \multicolumn{1}{c}{$5689$} & \multicolumn{1}{c}{$2774$} & 
\multicolumn{1}{c}{$2915$} \\ 
GRM & \multicolumn{1}{c}{$5709$} & \multicolumn{1}{c}{$2783$} & 
\multicolumn{1}{c}{$2925$} \\ \hline
\end{tabular}%
\caption{For various IRT models, the overall mean-squared predictive error (D), the goodness of fit (GF), and
the penalty (Pen).}\label{Model Compare}%
\end{table}%

\bigskip \newpage

\begin{center}
\textbf{Figure Captions}
\end{center}

\bigskip

\textbf{Figure 1}. \ For the simulated data, the marginal posterior mean
density estimates of the rating category thresholds, for two items.

\bigskip

\textbf{Figure 2}. \ For three items of the Verbal Aggression questionnaire,
trace plots of the MCMC posterior samples of the threshold parameters.

\bigskip

\textbf{Figure 3}. \ For six examinees of the Verbal Aggression
questionnaire, trace plots of the MCMC posterior samples of the ability
parameters.

\bigskip

\textbf{Figure 4}. \ For three items of the Verbal Aggression questionnaire,
the marginal posterior mean density estimate of the rating category
thresholds.

\bigskip

\textbf{Figure 5}. \ Median, interquartile, and 95-percentile range of the
marginal posterior distribution of the neighborhood location ($\gamma $) and
size ($\psi $), for each item of the Verbal Aggression questionnaire.\newline

\endgroup\medskip

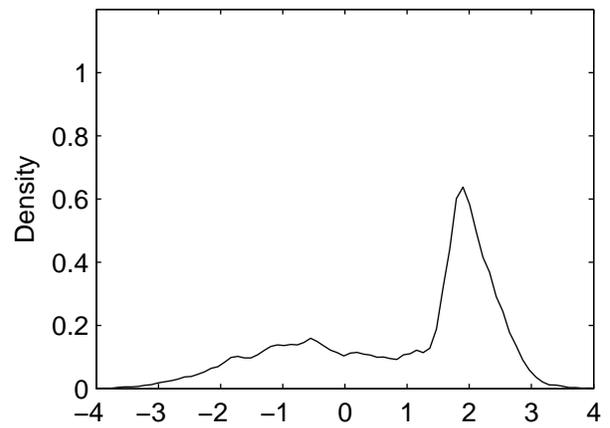
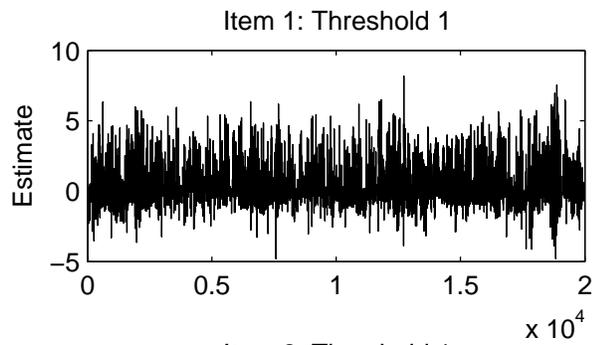
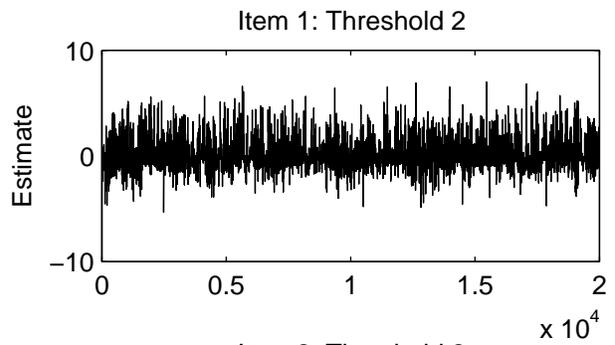
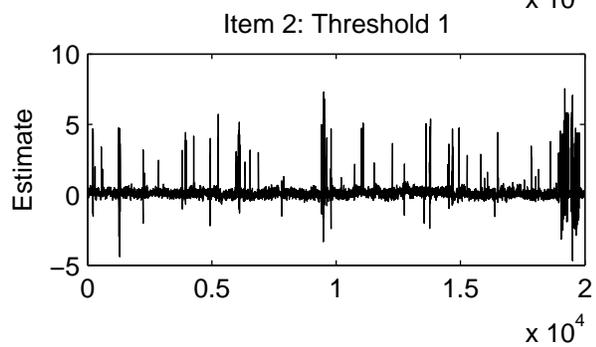
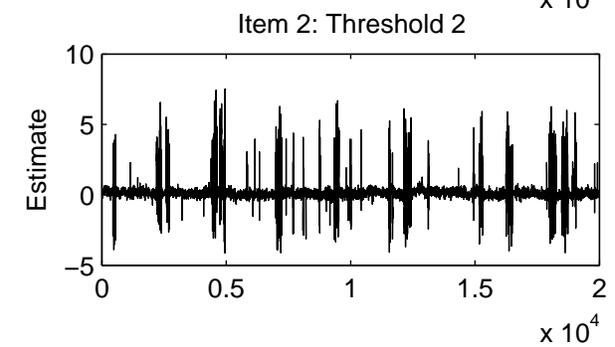
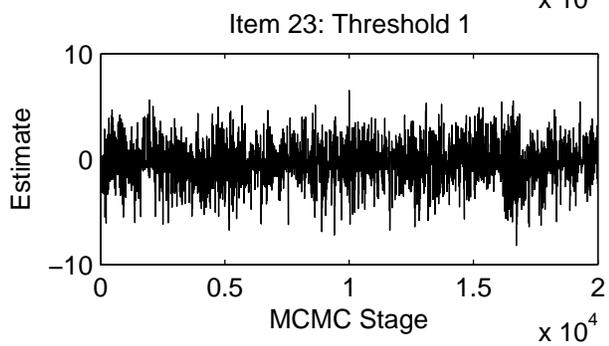
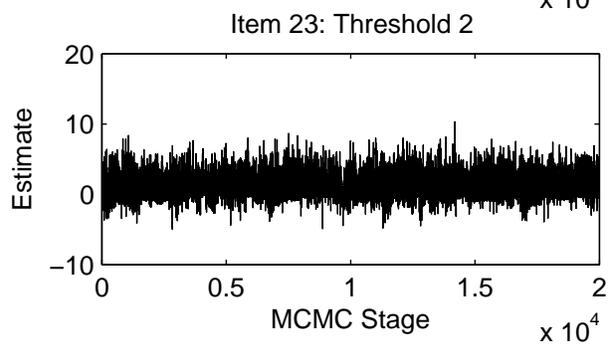
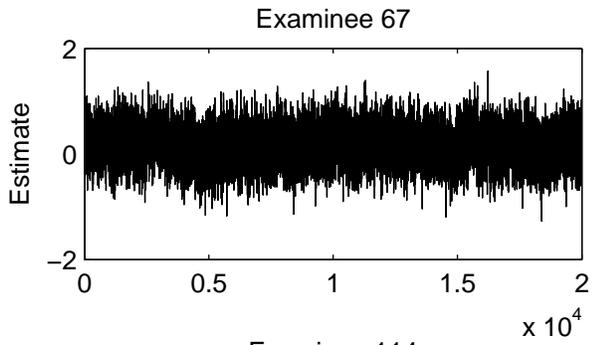
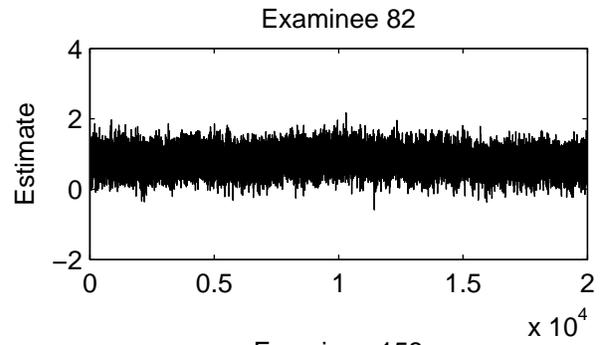
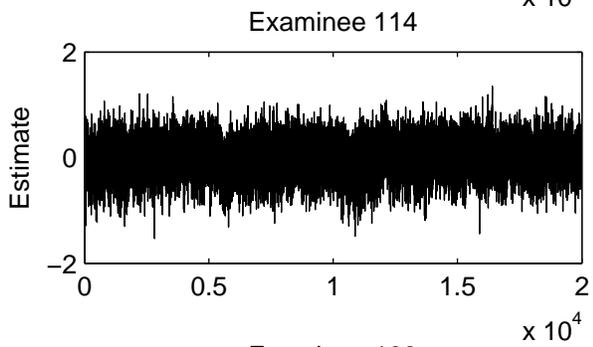
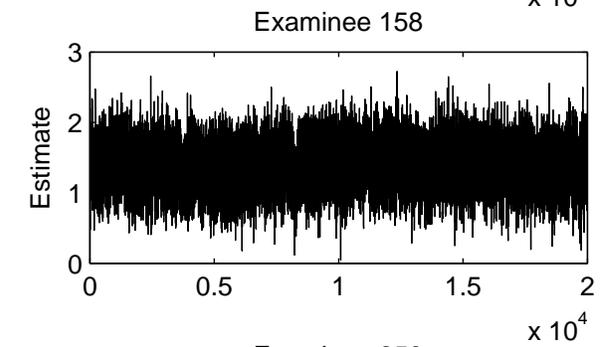
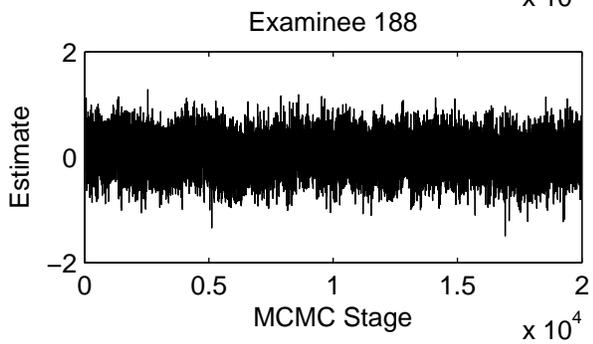
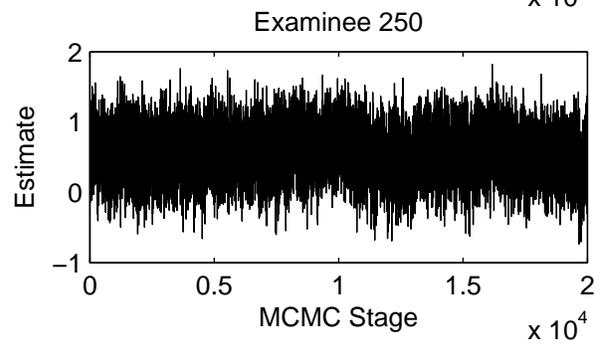
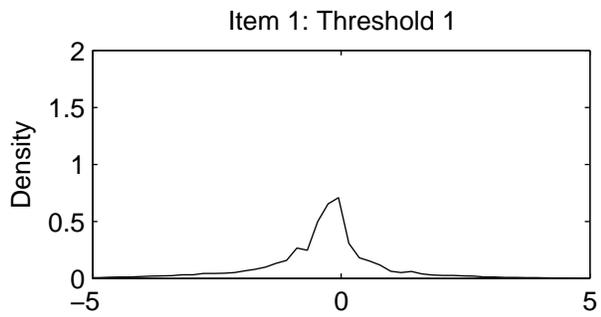
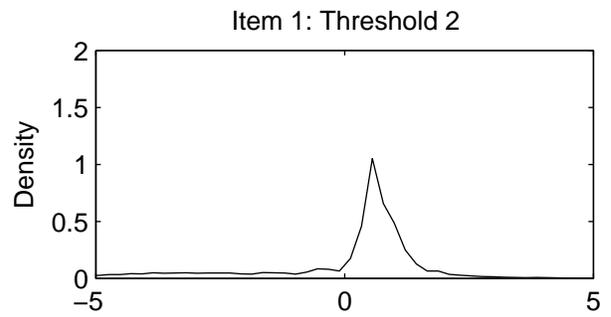
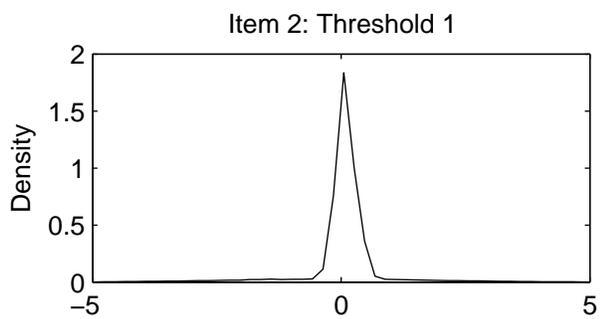
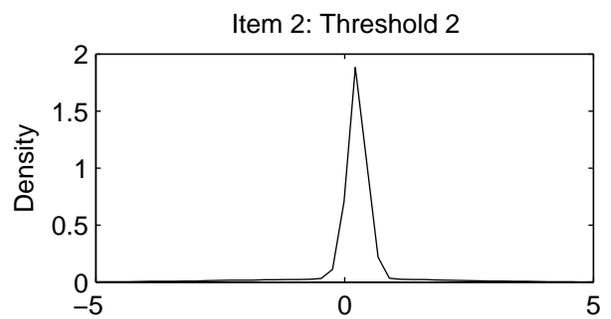
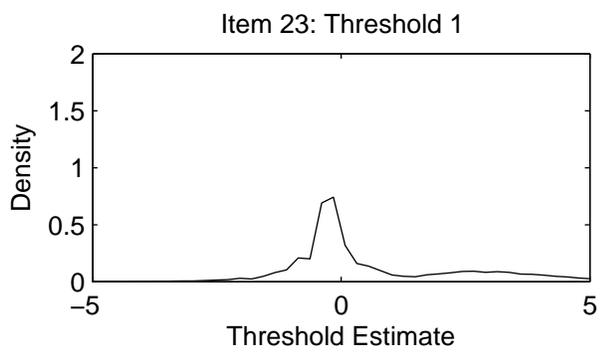
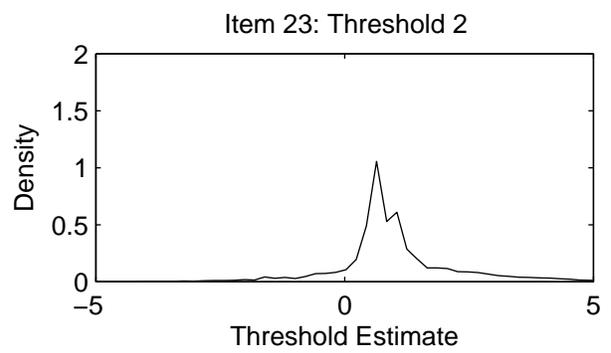
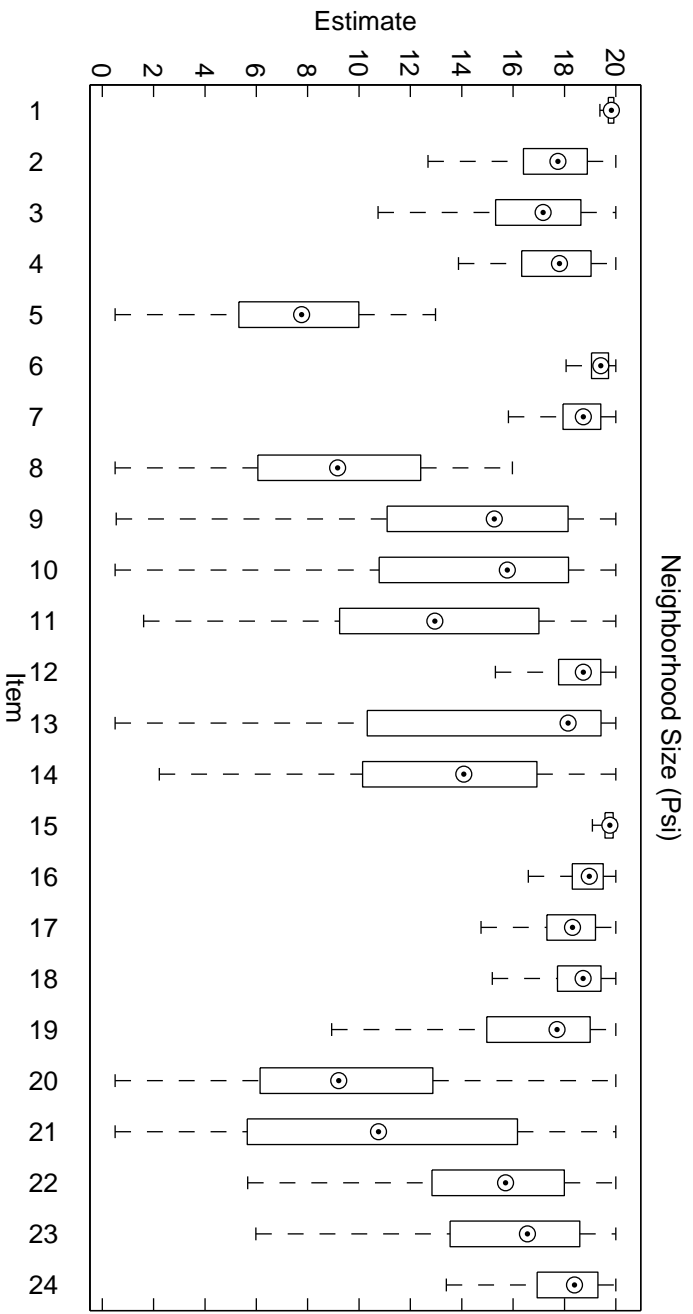
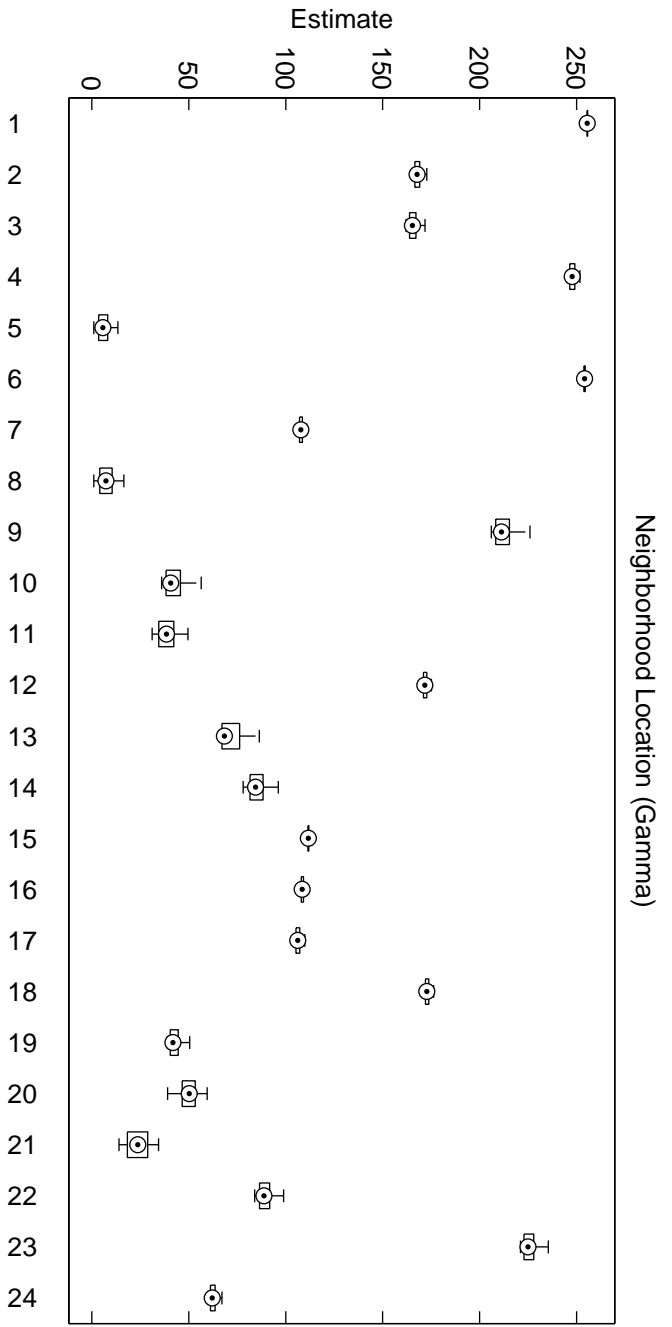
\end{document}